\documentstyle[12pt]{article}
\def\tilde{\widetilde}
\def\ta{U^{1}}
\def\tb{U^{2}}
\def\ga{G^{1}_{+}(v)}
\def\gb{G^{2}_{+}(v)}
\def\gaa{G^{1}_{-}(v)}
\def\gbb{G^{2}_{-}(v)}
\def\La{L^{1}_{+}}
\def\Lb{L^{2}_{+}}
\def\laa{L^{1}_{-}}
\def\lbb{L^{2}_{-}}
\def\Lc{L^{1}_{\pm}}

\def\mc{\tilde L^{1}_{\pm}}
\def\r{r_{+}}
\def\ra{r_{-}}
\def\rb{r_{\pm}}

\def\tr{\,{\rm tr}\,}

\def\a{\alpha}
\def\b{\beta}

\def\G{Gauss-law constraints }
\def\ui{U({\bf n};i)}

\def\n{{\bf n}}

\def\be{\begin{equation}}
\def\ee{\end{equation}}
\def\bea{\begin{eqnarray}}
\def\eea{\end{eqnarray}}
\def\Y{Yang-Mills }
\def\e{\epsilon}
\def\a{\alpha}
\def\h{the Heisenberg double }
\def\g{\gamma}
\vspace{5cm}

\title{  \hfill{LMU-TPW 95-3} \\ \hfill{hep-th/9502...  }
\\ Hamiltonian lattice gauge models and the Heisenberg double}
\vspace{.7cm}
\author{ \mbox{}
\\ S.A.Frolov\thanks{Alexander von Humboldt fellow}
\mbox{} \\ \vspace{0.4cm} Section Physik, Munich University
\vspace{-0.5cm} \mbox{} \\ Theresienstr.37, 80333 Munich, Germany
\thanks{Permanent address:\ Steklov Mathematical Institute, Vavilov st.42,
GSP-1, 117966 Moscow,
RUSSIA}
\mbox{} \\
\date{}}

\begin{document}
\maketitle
\vspace{3.5cm}
\begin{abstract}
Hamiltonian lattice gauge models based on the assignment of the Heisenberg
double of a Lie group to
each link of the lattice are constructed in arbitrary space-time dimensions. It
is shown that the
corresponding generalization of the gauge-invariant Wilson line observables
requires to attach to
each vertex of the line a vertex operator which goes to the unity in the
continuum limit.
\end{abstract}

 \section{Introduction}

In a recent paper \cite{fr} I have considered Hamiltonian lattice Yang-Mills
theory, based on the
assignment of the Heisenberg double of a Lie group to each link of the lattice,
in (1+1)- and
(2+1)-dimensions. In the present paper having in mind possible applications of
the construction to
Chern-Simons models and gravity I discuss an analogous formulation of lattice
gauge models on
an arbitrary lattice or graph. In the case of the regular hyper-cubic space
lattice the gauge models
proposed are lattice-regularized \Y models in (d+1)-dimensions.

As is well known there are two  possible ways of lattice regularization of
gauge
theories.
In the approach of Wilson \cite{w} one considers Euclidean formulation of a
model and discretizes
all space-time, thus replacing the model by some
statistical mechanics model. In the Hamiltonian approach of Kogut and Susskind
\cite{ks} one
considers a model in the Minkowskian space-time  and introduces only  space
lattice remaining
the time direction continuous. Then one places on each link of the lattice the
cotangent bundle of
a Lie group and on each vertex lattice \G which are first-class constraints and
generate gauge
transformations and finally one finds a gauge-invariant lattice Hamiltonian.
Thus in the Hamiltonian
 formulation  the \Y theory is replaced by some classical mechanics model with
first-class
 constraints, the phase space of the model being the direct product of the
cotangent bundle over
 all links of the lattice: $ \prod_{links} T^{*}G$.

However one could ask oneself whether it is possible to place on each link
another phase space
and on each vertex other \G and to get another lattice model which can be
reduced to the continuous
 one in the continuum limit. In \cite{fr} I have shown that such a possibility
does exist and is
based on a phase space which is called \h $D_{+}^{\g}$ of a Lie group and is
one
of the basic
objects in the theory of Poisson-Lie groups \cite{d1,s1}. Only (1+1)- and
(2+1)-dimensional \Y models were
considered in the paper. In the present paper a generalization of the
consideration to the \Y
theory in any space-time dimension is found.

The plan of the paper is as follows. In the second section we remind some
simple
results from the
theory of \h and introduce the notations used in the paper. In the third
section
we formulate a
classical mechanics lattice gauge model with first-class constraints on
arbitrary lattice or graph,
the phase space of the model being the direct product of \h over all links:
$\prod_{links} D_{+}^{\g}$. Then we show that in the case of the regular
hyper-cubic space lattice
in $d$-dimensions the model constructed is lattice-regularized
(d+1)-dimensional
\Y theory. In
Conclusion we discuss unsolved problems and perspectives.

\section{Heisenberg double }

In this section we remind some simple results from the theory of \h  and fix
notations. More
detailed discussion of the subject can be found in
refs.\cite{d1,s1,sr,af1,s2,am}.

Let $G$ be a matrix algebraic group and $D=G\times G$. For definiteness we
consider the case of
the $SL(N)$ group. Almost all elements $(x,y) \in D$ can be presented in two
equivalent forms  as
follows

\bea
(x,y)&=&(U,U)^{-1}(L_{+},L_{-})=(U^{-1}L_{+},U^{-1}L_{-}) \nonumber\\
&=&(\tilde L_{+}, \tilde L_{-})^{-1}(\tilde U,\tilde U)=
(\tilde L_{+}^{-1} \tilde U,\tilde L_{-}^{-1}\tilde U)
\label{2.1}
\eea
where $U, \tilde U \in G$, the matrices $L_{+}, \tilde L_{+}$ and $L_{-},\tilde
L_{-} $ are upper-
and lower-triangular, their diagonal parts $l_{+}, \tilde l_{+}$ and
$l_{-},\tilde l_{-}$ being
inverse to each other: $l_{+}l_{-}= \tilde l_{+}\tilde l_{-}=1$.

Let all of the matrices be in the fundamental representation  $V$ of the group
$G$ ($N\times
 N$ matrices for the $SL(N)$ group). Then the algebra of functions on the group
$D$ is generated by
 the matrix elements $x_{ij}$ and $y_{ij}$. The matrices $L_{\pm}$ and $U$ or
$\tilde L_{\pm}$ and
 $\tilde U$ can be considered as almost everywhere regular functions of $x$ and
$y$. Therefore, the
 matrix elements $L_{\pm ij}$ and
$U_{ij}$ (or $\tilde L_{\pm ij}$ and $\tilde U_{ij}$) define another system of
generators of the
algebra $Fun D$. We define the Poisson structure on the group $D$ in terms of
the generators
$L_{\pm}$ and $U$ as follows \cite{af1,s2}
\be
\{\ta ,\tb \} =\g [\rb ,\ta\tb ]
\label{2.2}
\ee
\bea
&&\{\La ,\Lb \} =\g [\rb ,\La\Lb ] \nonumber\\
&&\{\laa ,\lbb \} =\g [\rb ,\laa\lbb] \nonumber\\
&&\{\La ,\lbb \} =\g [\r ,\La\lbb]
\label{2.3}
\eea
\bea
&&\{\La ,\tb \} =\g \r \La\tb \nonumber\\
&&\{\laa ,\tb\} =\g \ra \laa\tb
\label{2.4}
\eea
Here $\g$ is an arbitrary complex parameter, $\rb$ are classical $r$-matrices
which satisfy the
classical Yang-Baxter equation and the following relations
\be
[r^{12} ,r^{13} ]+[r^{12} ,r^{23} ]+[r^{13} ,r^{23} ]=0
\label{2.5}
\ee
\be
\ra =-P\r P, \qquad \r -\ra =C
\label{2.6}
\ee
where $P$ is a permutation in the tensor product $V\otimes V$ ($Pa\otimes
b=b\otimes a$) and the
matrix $C$ is the tensor Casimir operator of the Lie algebra of the group $G$.
For the $SL(N)$ group
 the solution of eqs.(2.5-2.6) looks as follows
\bea
\r &=& \sum_{i=1}^{N-1} h_{i}\otimes  h_{i} +2\sum_{i<j}^{N} e_{ij}\otimes
e_{ji} \nonumber\\
&=&-\frac {1}{N} I + \sum_{i=1}^{N} e_{ii}\otimes e_{ii}  + 2\sum_{i<j}^{N}
e_{ij}\otimes e_{ji}
\label{2.7}
\eea
where $(e_{ij})_{kl} =\delta_{ik}\delta_{jl}$ and $h_{i}$ form an orthonomal
basis of the Cartan
subalgebra of the $SL(N)$ group: $\sqrt{\, i(i+1)} h_{i}=\sum_{k=1}^{i}
e_{kk}-ie_{i+1,i+1}$.

\noindent In eqs.(2.2-2.6) we use the standard notations from the theory of
quantum groups
\cite{d2,frt}:  for any matrix $A$ acting in some space $V$ one can
construct two matrices $A^{1}=A\otimes id $ and $A^{2}=id\otimes A $ acting in
the space
$V\otimes V$, and for any matrix $r=\sum_{a} r_{1}(a)\otimes r_{2}(a)$ acting
in
the space
$V\otimes V$ one can construct matrices $r^{12}=\sum_{a} r_{1}(a)\otimes
r_{2}(a)\otimes id$,
$r^{13}=\sum_{a} r_{1}(a)\otimes id\otimes r_{2}(a)$ and
$r^{23}=\sum_{a} id\otimes r_{1}(a)\otimes r_{2}(a)$ acting in the space
$V\otimes V\otimes V$.

The group $D$ endowed with the Poisson structure (\ref{2.2}-\ref{2.4}) is
called
the Heisenberg
double $D_{+}^{\g}$ of the group $G$.  It is not difficult to show that the
matrices
$\tilde L_{\pm}$ and $\tilde U$ have the same Poisson structure (2.2-2.4) and
we
shall need the
Poisson brackets of $L_{\pm}$, $U$ and $\tilde L_{\pm}$, $\tilde U$ \cite{af2}
\bea
\{L_{\a}^{1},\tilde L_{\b}^{2}\} &=&0 \qquad for \quad any \quad \a ,\b =+,-
\nonumber \\
\{\mc ,\tb\} &=&-\g\mc \tb \rb \nonumber \\
\{\Lc ,\tilde\tb\} &=&-\g\Lc \tilde\tb \rb \nonumber \\
\{\ta ,\tilde\tb\} &=&0
\label{2.8}
\eea
The cotangent bundle of the group $G$ can be considered as a limiting case of
\h
. Namely, in the
limit $\g \to 0$ and  $L_{\pm} \to 1+\g E_{\pm}$ the Poisson structure of \h
coincides with the
canonical Poisson structure of the cotangent bundle $T^{*}G$.

Up to now we considered \h of a complex Lie group.   One can single out some
real forms by means of
the following (anti)-involutions:

\noindent 1. $SU(N)$ form for imaginary $\g$
\be
U^{*}=U^{-1},\quad L_{+}^{*}=L_{-}^{-1},\quad L_{-}^{*}=L_{+}^{-1}
\label{2.9}
\ee

\noindent 2. $SU(N)$ form for real $\g$ \cite{sch,af2}
\be
U^{*}=\tilde U,\quad  L_{+}^{*}=L_{-},\quad L_{-}^{*}=L_{+}
\label{2.10}
\ee

\noindent 3. $SL(N)$ form for real $\g$
\be
U^{*}= U,\quad L_{+}^{*}=L_{+},\quad L_{-}^{*}=L_{-}
\label{2.11}
\ee
It can be easily checked that these (anti)-involutions are compatible with the
Poisson structure
(2.2-2.4).

\section{Hamiltonian lattice Yang-Mills theory}

In this section we consider Hamiltonian lattice gauge models based on the
assignment of \h to each
link of the lattice. We begin with the case of an arbitrary graph (regular
hyper-cubic lattice,
triangulation of a surface, the Bruhat-Tits tree, simplicial complexes and so
on) which is described
by a set of vertices and a set of links. Each link is thought of as either a
path connecting two
vertices $v_{1}$ and $v_{2}$ or a closed path with a marked vertex (tadpole).
Two vertices can be
connected by any finite number of links. Such a graph is certainly just an
arbitrary connected
Feynman diagram.

Let us now consider some vicinity of a vertex $v$ which does not contain other
vertices and closed
paths. Let us denote the paths which go from the vertex $v$ by
$l_{1}(v)$,...,$l_{N_{v}}(v)$. We
call such a path as a vertex path. $N_{v}$ is a common number of the paths and
if there is no closed
path for the vertex $v$ then $N_{v}$ coincides with the number of links going
from $v$ to some other
vertices of the lattice. With each vertex path $l_{i}(v)$ one associates a
field
taking values in \h
$D_{+}^{\g}$. This field is described by matrices $U(l_{i}(v))$,
$L_{+}(l_{i}(v))$ and
$L_{-}(l_{i}(v))$ with the Poisson structure (2.2-2.4) and fields corresponding
to different paths
have vanishing Poisson brackets.

Let us now attach to the vertex $v$ the following Gauss-law constraints
\cite{fr}
\be
G_{\pm}(v)= L_{\pm}(l_{1}(v))L_{\pm}(l_{2}(v))\cdots L_{\pm}(l_{N_{v}}(v))=1
\label{3.1}
\ee
One can easily check that these constraints have the following  Poisson
brackets
\bea
&&\{\ga ,\gb \} =\g [\rb ,\ga\gb ]  \nonumber\\
&&\{\gaa ,\gbb \} =\g [\rb ,\gaa\gbb]  \nonumber\\
&&\{\ga ,\gbb \} =\g [\r ,\ga\gbb]
\label{3.2}
\eea
We see that these Poisson brackets vanish on the constraints surface
$G_{\pm}=1$
and therefore
they are first-class constraints. Thus one can consider gauge transformations
which are
generated by these constraints. Namely, for any function $X$ of the matrices
$U(l_{i}(v))$,
$L_{\pm}(l_{i}(v))$ an infinitesimal gauge transformation looks as follows
\be
\delta X=\{ X,\tr (G_{+}(v)\xi_{+} (v) +G_{-}(v)\xi_{-} (v))\}
\label{3.3}
\ee
where $\xi_{\pm}$ are gauge parameters which do not depend on $U$ and
$L_{\pm}$.

\noindent Using the Poisson brackets (2.4) for $U$ and $L_{\pm}$ one can easily
verify that these
constraints generate left gauge transformations of the field $U$
\be
U(l_{i}(v))\to g(l_{i}(v))U(l_{i}(v))
\label{3.4}
\ee
Let us note that $g(l_{i}(v))$ depends not only on the vertex $v$ but on the
vertex path $l_{i}(v)$
and the fields $L_{\pm}$ as well. But the gauge transformations for different
vertex paths are
certainly not independent.

A remarkable feature of these
constraints is that they form  the same quadratic Poisson algebra as
the matrices $L_{\pm}$ do. In the limit $\g \to 0$, $L_{\pm}\to 1+\g E_{\pm}$
 one gets the following Gauss-law constraints
\be
C_{\pm}(v)= \sum_{i=1}^{N_{v}} E_{\pm}(l_{i}(v))=0
\label{3.5}
\ee
which form the Lie algebra and were used by Kogut and Suskind \cite{ks} (more
exactly they used the
constraints $C(v)=C_{+}(v)-C_{-}(v)$). Thus the usual lattice gauge theory can
be thought of as a
particular case of the models under consideration.

Let us remark that the choice of the constraints $G_{\pm}(v)$ is not unique.
One
can use any
constraint of the form
\be
G_{\pm}(v, \sigma )= L_{\pm}(l_{\sigma_{1}}(v))L_{\pm}(l_{\sigma_{2}}(v))\cdots
L_{\pm}(l_{\sigma_{N_{v}}}(v))=1
\label{3.6}
\ee
where $\sigma\in Symm(N_{v})$ is some permutation of 1,2,...,$N_{v}$.

\noindent These constraints form the same Poisson algebra (\ref{3.2}) and in
the
limit $\g \to 0$
coincide with $C_{\pm}(v)$. However at finite $\g$ only that constraints, which
differ from each
other by a cyclic permutation, are equivalent. Thus with each vertex one can in
principle associate
$(N_{v}-1)!$ nonequivalent constraints.

Repeating the same procedure for all of the vertices one gets the phase space
which is the direct
product of \h over all of the vertex paths and a set of the \G attached to the
vertices. The \G
corresponding to different vertices have vanishing Poisson brackets. Taking
into
account that for
each link there are two vertex paths one sees that one has placed on each link
two different
Heisenberg doubles. However one can impose on the fields attached to one link
the following
constraints
\be
U^{-1}(1)L_{\pm}(1)=L_{\pm}^{-1}(2)U(2)
\label{3.7}
\ee
Comparing eq.(\ref{3.7}) with eq.(2.1) one concludes that the fields $U(1)$,
$L_{\pm}(1)$ and
$U(2)$, $L_{\pm}(2)$ are just different coordinates on the same Heisenberg
double. Thus the phase
space of the model is the direct product of \h over all links: $\prod_{links}
D_{+}^{\g}$.

To the moment we have described the phase space of the model and imposed the \G
. The next problem
is to find gauge-invariant functions on the phase space and to form from them a
Hamiltonian. The
simplest gauge-invariant functions can be obtained using a well-known theorem
from the theory of
Poisson-Lie groups \cite{d1,s1,s2} which states that generators of the ring of
the Casimir functions
of the Poisson algebra (2.3) have the following form
\be
h_{k}=\tr (L_{+}L_{-}^{-1})^{k}=\tr L^{k}
\label{3.8}
\ee
Due to the fact that the \G (\ref{3.1}) depend only on $L_{\pm}$ these
functions
are gauge-invariant.
The functions (\ref{3.8}) are not the only ones gauge-invariant. Just as in the
case of the usual
lattice
gauge theory one can construct Wilson line observables. So let us consider a
loop with is formed by
the oriented links $l_{1},l_{2},...,l_{n}$. Each oriented link which goes from
a
vertex $v$ to a
vertex $u$ (it may be the same vertex) can be denoted by two vertex paths as
follows
\be
l(v,u)=l_{i}(v)l_{j}(u)
\label{3.9}
\ee
where $i$ and $j$ are the numbers of the vertex paths (we have numbered all
vertex paths).
\noindent Thus the loop is described by $2n$ vertex paths as follows
\be
l_{1},l_{2},...,l_{n}=l_{i_{1}}(v_{1})l_{j_{2}}(v_{2}),
l_{i_{2}}(v_{2})l_{j_{3}}(v_{3}),...,
l_{i_{n}}(v_{n})l_{j_{1}}(v_{1})
\label{3.10}
\ee
i.e. we denote the link $l_{\alpha}$ going from $v_{\alpha}$ to $v_{\alpha +1}$
as
$l_{i_{\alpha}}(v_{\alpha})l_{j_{\alpha +1}}(v_{\alpha +1})$ and
$i_{\alpha},j_{\alpha}=1,2,...,N_{v_{\alpha}}$.

\noindent In the usual lattice gauge theory the gauge-invariant Wilson line
observable corresponding
to the loop is of the form
\be
W(l_{1}\cdots l_{n})=\tr U(l_{i_{1}}(v_{1}))U(l_{i_{2}}(v_{2}))\cdots
U(l_{i_{n}}(v_{n}))
\label{3.11}
\ee
We are looking for a similar expression for the Wilson line observable which
coincides with
(\ref{3.11})
in the limit $\g \to 0$. It appears that to get a corresponding generalization
one should attach to
every vertex $v_{\alpha}$ some vertex operator
$V(v_{\alpha},j_{\alpha},i_{\alpha})$ which depends
on the vertex paths $l_{i_{\alpha}}(v_{\alpha})$ and
$l_{j_{\alpha}}(v_{\alpha})$. Then the
gauge-invariant Wilson line observable for a loop without tadpoles looks as
follows
\be
W(l_{1}\cdots l_{n})=\tr U(i_{1})V(j_{2},i_{2})U(i_{2})V(j_{3},i_{3})\cdots
U(i_{n})V(j_{1},i_{1})
\label{3.12}
\ee
where we use short notations
$$ U(l_{i_{\alpha}}(v_{\alpha}))=U(i_{\alpha}) $$
$$ V(v_{\alpha},j_{\alpha},i_{\alpha})=V(j_{\alpha},i_{\alpha}) $$

The choice of the vertex operator $V(j,i)$ is not unique. The simplest vertex
operators have the
form
\be
V(j,i)=\left\{ \begin{array}{cc} L_{\e_{j+1}}(j+1)\cdots L_{\e_{i-1}}(i-1) &
if\quad j<i-1 \\
1 & if\quad j=i-1 \\
L_{\mu_{j}}^{-1}(j)\cdots L_{\mu_{i}}^{-1}(i) & if\quad j\ge i \end{array}
\right.
\label{3.13}
\ee
where $\e_{k} ,\mu_{l} =\pm$.

\noindent To prove eq.(\ref{3.12}) it is enough to show that the combination
\be
U(l_{k}(u))V(v,j,i)U(l_{i}(v))=\tilde U(l_{j}(v))V(v,j,i)U(l_{i}(v))
\label{3.14}
\ee
is gauge-invariant under the transformations generated by the constraints
$G_{\pm}(v)$. It is not
difficult to do by using formulas (2.2-2.4, 2.8) for the Poisson brackets of
$U$, $\tilde U$ and
$L_{\pm}$. A Hamiltonian of the model can be now written as some combination of
the gauge-invariant
functions (\ref{3.8}) and (\ref{3.12}).

Thus we have described the phase space, the constraints and gauge-invariant
observables which can be
used as Hamiltonians of the models and now we are passing to a particular model
on the regular
hyper-cubic space lattice. We shall show that this model is the
lattice-regularized \Y theory.

So let us consider the regular hyper-cubic space lattice in $d$-dimensions. In
this case an arbitrary
vertex of the lattice can be denoted by a vector ${\bf n}=(n_{1},...,n_{d})$
with
integers $n_{i}$. We choose some orientation of the lattice and denote
the orthonormal lattice vectors which define the orientation as  ${\bf e}_{i}$,
$i=1,...,d$ and
introduce the notation ${\bf e}_{-i}=-{\bf e}_{i}$. With each link one can
associate a positive link
$({\bf n},{\bf e}_{i})$ and a negative link $({\bf n}+{\bf e}_{i},-{\bf
e}_{i})$. The vertex paths
of the vertex ${\bf n}$ are therefore oriented links $({\bf n},{\bf
e}_{\alpha})$, $\alpha = \pm
1,...,\pm d$. We place on each vertex path $({\bf n},{\bf e}_{\alpha})$ (or
oriented link) \h which
is described by the fields $U(\n ,\a), L_{\pm}(\n ,\a)$. Due to eq.(\ref{3.7})
we have the following
relations
\bea
U({\bf n}+{\bf e}_{i},-i)&=&\tilde U(\n ,i) \nonumber\\
L_{\pm}({\bf n}+{\bf e}_{i},-i)&=&\tilde L_{\pm}(\n ,i)
\label{3.15}
\eea
Let us note that in the limit $\g \to 0$, $L_{\pm}\to 1+\g E_{\pm}$ one gets
the
usual equations
$U(l^{-1})=U^{-1}(l)$ and $E(l^{-1})=-U^{-1}(l)E(l)U(l)$  for any oriented link
$l$.

There are $(2d-1)!$ nonequivalent choices of the constraints (\ref{3.1}). In
the
paper we use the
following \G
\bea
G_{\pm}(\n)&=&L_{\pm}(\n ,-1) L_{\pm}(\n ,1)L_{\pm}(\n ,-2)L_{\pm}(\n ,2)\cdots
L_{\pm}(\n ,-d)
L_{\pm}(\n ,d) \nonumber\\
&=&\tilde L_{\pm}(\n -{\bf e}_{1},1)L_{\pm}(\n ,1)\cdots L_{\pm}(\n -{\bf
e}_{d},d)L_{\pm}(\n ,d)=1
\label{3.16}
\eea
We see from this expression that it is natural to introduce the notation
$G_{\pm}(\n ,i)=L_{\pm}(\n ,-i)L_{\pm}(\n ,i)$. One can easily verify that the
(anti)-involutions
(2.9) and (2.11) are compatible with the constraints. A Hamiltonian of the
model
can be written in
the following form (which is certainly not unique)
\be
H=\frac {e^{2}}{2\g^2}a^{2-d} \sum_{links}\tr (L^{2}(l)-1) +
\frac {a^{d-4}}{2e^{2}}\sum_{plaqettes}\big(W(\Box) + W^{*}(\Box)\big)
\label{3.17}
\ee
where the summation is taken over all positive and negative links and over all
plaqettes, $e$ is the
coupling constant and $a$ is the lattice length.
The Wilson term $W(\Box) $ is determined by eqs.(\ref{3.11},\ref{3.13}) and is
equal to
\bea
W(\Box _{ij})&=& \tr \ui V(\n +{\bf e}_{i};-i,j)U(\n +{\bf e}_{i};j)
V(\n +{\bf e}_{i}+{\bf e}_{j};-j,-i) \nonumber \\
&&U(\n +{\bf e}_{i}+{\bf e}_{j};-i)V(\n +{\bf e}_{j};i,-j) U(\n +{\bf
e}_{j};-j)V(\n ;j,i)
\label{3.18}
\eea
This formula can be simplified if one uses the following equation expressing
$U(\n +{\bf e}_{i};-i)$
through $\ui$, $L_{\pm}(\n ,i)$ and $\tilde L_{\pm}(\n ,i)$
\be
U(\n +{\bf e}_{i};-i)= \tilde L_{\pm}(\n ,i)U^{-1}(\n ,i)L_{\pm}(\n ,i)=
L_{\pm}(\n+{\bf e}_{i};-i)U^{-1}(\n ,i)L_{\pm}(\n ,i)
\label{3.19}
\ee
Using this equation and eq.(\ref{3.13}) for the vertex operators one gets for
$W(\Box)$
\bea
W(\Box _{ij})&=& \tr \ui V_{ij}(\n +{\bf e}_{i})U(\n +{\bf e}_{i};j)
V_{ji}(\n +{\bf e}_{i}+{\bf e}_{j}) \nonumber \\
&&U^{-1}(\n +{\bf e}_{j};i)V_{ij}(\n +{\bf e}_{j})U^{-1}(\n ;j)V_{ji}(\n )
\label{3.20}
\eea
where
\be
V_{ij}(\n)=\left\{ \begin{array}{cc} L_{\e_{i}}(\n ,i)G_{\e_{i+1}}(\n
,i+1)\cdots
G_{\e_{j-1}}(\n ,j-1)L_{\e_{j}}(\n ,-j) & if \quad i<j \\
1 & if \quad i=j \\
(L_{\mu_{j}}(\n ,j)G_{\mu_{j+1}}(\n ,j+1)\cdots G_{\mu_{i-1}}(\n
,i-1)L_{\mu_{i}}(\n ,-i))^{-1}
 & if\quad  i>j \end{array} \right.
\label{3.21}
\ee
Let us mention that the Wilson line observable corresponding to an arbitrary
loop formed by oriented
links $l_{1},l_{2},...,l_{n}$ has the same form
\be
W(l_{1}l_{2}...l_{n})=\tr U(l_{1})V_{l_{1}l_{2}}U(l_{2})V_{l_{2}l_{3}}\cdots
U(l_{n})V_{l_{n}l_{1}}
\label{3.22}
\ee
where one should take $U(l)=\ui $ for any positive link $l=({\bf n},{\bf
e}_{i})$
and $U(l^{-1})=U^{-1}(\n ,i)$ for any negative link $l^{-1}=({\bf n}+{\bf
e}_{i},-{\bf e}_{i})$ and
the vertex operator $V_{l_{k}l_{k+1}}$ is given by eq.(\ref{3.21}). Let us
stress that in
eq.(\ref{3.22}) one has to use $U^{-1}(\n ,i)$ for any negative link but not
$\tilde U(\n ,i)$.

\noindent One can choose for example the positive sign for all of $\e_{k}$ and
negative sign for all
of $\mu_{l}$. Then the vertex operator $V_{ij}(\n)$ has the following
transformation law with
respect to the anti-involution (2.9) which singles out the $SU(N)$ real form
\be
V_{ij}^{*}(\n)=V_{ji}(\n)
\label{3.23}
\ee
The formula (\ref{3.23}) ensures that the Wilson line observable
$W(l_{1}l_{2}...l_{n})$ is
complex- conjugated to $W(l_{n}l_{n-1}...l_{1})$.

Now taking into account that in the limit $\g \to 0$, $L_{\pm}\to 1+\g E_{\pm}$
the vertex operator
$V_{ij}$ goes to the unity one recovers the \G and the Hamiltonian of the usual
lattice gauge theory
\cite{ks}. Thus we have shown that in the case of the regular hyper-cubic space
lattice the model
proposed is just a lattice-regularized \Y theory.

\section{Conclusion}

In this paper we considered the  Hamiltonian formulation of classical gauge
models on an arbitrary
lattice.  In this formulation we placed on each link
\h of a Lie group and attached \G to each vertex of the lattice. We have shown
that the models on
the regular hyper-cubic lattices correspond to lattice-regularized \Y theory.
We
discussed only the
case of the pure \Y theory. It would be very interesting to include fermions to
this construction.
It seems to be plausible that a proper description of fermions would lead to a
lattice version of
the Faddeev-Shatashvili-Mickelsson 2-cocycle \cite{f,fs,m}.

Another interesting problem is to classify all integrable lattice gauge models,
i.e. to find all
corresponding graphs and Hamiltonians.

We considered only classical theory and the next and most important problem is
to quantize the models.
There is no problem in quantizing the Poisson structure of the Heisenberg
double. One just gets the
quantized algebra of functions on \h \cite{af1,s2,af2} which includes as
subalgebras the algebra of
functions on the quantum group $Fun_{q}(G)$ and the quantized universal
enveloping
algebra $U_{q}(\cal G)$, where $\cal G$ is the Lie algebra of the group $G$.
The
classical
$r$-matrices $\rb$ are to be replaced by the $R$-matrices
$R_{\pm}(q)=1+i\hbar\g\rb +\cdots$,
where $q={\rm e}^{i\hbar\g}$. Thus a real $\g$ corresponds to $q$ lying on the
unit circle of the
complex plane and an imaginary $\g$ corresponds to a real $q$. It is of no
problem to check that
in quantum theory the \G are first-class constraints and commute with the
quantum
Wilson line observables and therefore with the quantum Hamiltonian. Let us note
that $q$ has a
nonpolynomial dependence on the Planck constant $\hbar$ and thus already "tree"
correlation functions
of the models will have a nonpolynomial dependence on $\hbar$ as well. It seems
to be an indication that
correlation functions of the models correspond to a summation over
infinitely-many number of the
usual Feynman diagrams. It is not excluded that the parameter $\g$ plays the
role of an infrared
cut-off. Due to the fact that there is the additional parameter $\g$ for the
models one may expect
that these models have more rich phase structure than the usual lattice gauge
theory.

Let us finally notice that $q$-deformed lattice gauge theory was considered in
refs.\cite{fock,b,ags,br} in connection with  the Chern-Simons theory.

{\bf Acknowledgements:} I am grateful to
Professor J.Wess for kind hospitality and the Alexander von Humboldt
Foundation for the support. This work has been supported in part by
ISF-grant MNB000 and by the Russian Basic Research Fund under
grant number 94-01-00300a.

\end{document}